\documentclass[twocolumn,showpacs,amssymb,aps,prc,nofootinbib,floatfix,superscriptaddress]{revtex4-1}
\usepackage{amsmath,graphicx,color,ulem}
\usepackage{hyperref}

\begin{document}
\title{Explaining higher-order correlations between elliptic and triangular flow}

\author{Mubarak Alqahtani}
\affiliation{Department of Physics, College of Science, Imam Abdulrahman Bin Faisal University, Dammam 31441, Saudi Arabia  }
\affiliation{Basic and Applied Scientific Research Center, Imam Abdulrahman Bin Faisal University, Dammam 31441 Saudi Arabia}
\author{Jean-Yves Ollitrault}
\affiliation{Universit\'e Paris Saclay, CNRS, CEA, Institut de physique th\'eorique, 91191 Gif-sur-Yvette, France}

\begin{abstract} 
The ALICE and CMS Collaborations have analyzed a number of cumulants mixing elliptic flow ($v_2$) and triangular flow ($v_3$), involving up to $8$ particles, in Pb+Pb collisions at the LHC. 
We unravel an unexpected simplicity in these complex mathematical quantities for collisions at fixed impact parameter. 
We show that as one increases the order in $v_2$,  for a given order in $v_3$, the changes in the cumulants are solely determined by the mean elliptic flow in the reaction plane, which originates from the almond-shaped geometry of the overlap area between the colliding nuclei. 
We derive simple analytic relations between cumulants of different orders on this basis. 
These relations are in good agreement with recent data from the CMS Collaboration. 
We argue that agreement will be further improved if the analysis is repeated with a finer centrality binning. 
We make quantitative predictions for cumulants of order 10 which have not yet been analyzed. 
\end{abstract}
\maketitle

\section{Introduction}
\label{s:introduction}

Cumulants of multiparticle azimuthal correlations are routinely analyzed in nucleus-nucleus collisions~\cite{STAR:2002hbo,NA49:2003njx,ALICE:2010suc,ALICE:2011ab,CMS:2012zex,CMS:2013wjq,ATLAS:2014qxy,STAR:2015mki,ALICE:2018lao}.
They were originally introduced \cite{Borghini:2000sa,Borghini:2001vi} in order to separate elliptic flow in the direction of impact parameter, induced by pressure gradients within the almond-shaped quark-gluon plasma~\cite{Ollitrault:1992bk}, from other correlations of various origins (Bose-Einstein~\cite{Dinh:1999mn}, global momentum conservation \cite{Borghini:2000cm,Bzdak:2010fd}, resonance decays~\cite{Feng:2021pgf}, jets~\cite{Kovchegov:2002nf,Connors:2017ptx}), referred to as ``nonflow''  correlations, whose contribution decreases rapidly as the order of the cumulant (the number of correlated particles) increases. 
The perspective changed when it was realized that local fluctuations in the initial energy density contribute significantly to pressure gradients, generating  large event-by-event fluctuations of elliptic flow~\cite{Miller:2003kd,Andrade:2006yh,PHOBOS:2006dbo}, and a new phenomenon solely due to fluctuations,  triangular flow~\cite{Alver:2010gr}. 

This led to a paradigm shift. 
One now considers that flow is the dominant contribution to all cumulants, even the lowest-order ones (pair correlations), provided that there is a rapidity gap between the particles that are correlated~\cite{PHENIX:2003qra}.  
Thus, cumulants are solely determined by the probability distribution of flow fluctuations~\cite{ATLAS:2013xzf,ALICE:2018rtz}, which is used in order to constrain models of the initial energy density~\cite{Retinskaya:2013gca,Moreland:2014oya}. 
This turns out to be the most robust aspect of global theory-to-data comparison~\cite{JETSCAPE:2020mzn}.

In this paper, we unravel the information contained in cumulants of azimuthal correlations involving both elliptic flow ($v_2$) and triangular flow ($v_3$), which are called  ``mixed harmonic cumulants'' (MHC)~\cite{ALICE:2021adw,CMS:2025opi}. 
$v_2$ and $v_3$ are the largest Fourier harmonics of anisotropic flow~\cite{ALICE:2011ab}. 
Relativistic hydrodynamics predicts~\cite{Niemi:2012aj} that they are approximately proportional to the initial anisotropies $\varepsilon_2$ and $\varepsilon_3$, which are Fourier coefficients of the initial density profile. 
Therefore, cumulants involving $v_2$ and $v_3$ constrain models of the initial state more directly than higher harmonics, which are largely driven by nonlinear response terms~\cite{Borghini:2005kd,Gardim:2011xv}. 
The lowest-order MHC is a 4-particle cumulant introduced by Bilandzic~{\it et al.\/} and dubbed a ``symmetric cumulant'' (SC)~\cite{Bilandzic:2013kga}-, which has been measured in Pb+Pb collisions at the LHC~\cite{ALICE:2016kpq,CMS:2025opi,ATLAS:2019peb}. 
It represents the linear correlation between $v_2^2$ and $v_3^2$, which is negative except in ultra-central collisions. 
We have recently shown~\cite{Alqahtani:2025wan} that it is driven by the correlation between the elliptic flow in the reaction plane, $v_{2,x}$, and $v_3^2$.
Our goal is to extend this study to higher-order cumulants, of 6 and 8 particles, which have subsequently been measured~\cite{ALICE:2021adw,CMS:2025opi}. 

Throughout this paper, we assume that nonflow correlations are negligible, so that particles in each event are emitted independently according to an underlying probability distribution~\cite{Luzum:2011mm,Ollitrault:2023wjk}. 
Let $P(\varphi)$  denote the azimuthal dependence of this probability distribution. 
 The complex anisotropic flow $V_n$ of the event is defined as its Fourier coefficient of order $n$~\cite{Luzum:2013yya}:
\begin{equation}
  V_n\equiv \int_{0}^{2\pi}e^{in\varphi}P(\varphi)d\varphi. 
\label{defvn}
\end{equation}
The usual anisotropic flow $v_n$ is the modulus,  $v_n\equiv |V_n|$. 

In Sec.~\ref{s:expcumulants}, we define the cumulants of $v_2$ and $v_3$ using the formalism of generating functions and we compare data from several analyses in Pb+Pb collisions at the LHC. 
In Sec.~\ref{s:intrinsiccumulants}, we define similar cumulants but in a different coordinate frame, the ``intrinsic'' frame where the direction of impact parameter is fixed~\cite{Roubertie:2025qps,Alqahtani:2025wan}. 
In this frame, the magnitude of cumulants decreases as a function of the order, in a way which follows specific scaling rules, provided that all events have the same impact parameter. 
In Sec.~\ref{s:expvsintrinsic}, we express the experimental cumulants as a function of the cumulants in the intrinsic frame.  
In Sec.~\ref{s:ratios}, we derive relations between experimental cumulants of different orders, and compare them with LHC data.

\section{Cumulants of flow fluctuations in the laboratory frame}
\label{s:expcumulants}

\subsection{Definitions}

The quantities that can be measured experimentally are moments of the  joint distribution of $V_2$ and $V_3$~\cite{Bhalerao:2011yg,Bhalerao:2014xra}. 
The cumulants $MHC(v_2^{2m},v_3^{2q})$ analyzed by ALICE~\cite{ALICE:2021adw} and CMS~\cite{CMS:2025opi}, where $m$ and $q$ are positive integers, are combinations of these moments~\cite{Bilandzic:2010jr,Bilandzic:2013kga}.\footnote{The ALICE collaboration has subsequently analyzed a different set of ``cumulants'' which are different combination of moments~\cite{ALICE:2023lwx}. They are not derived from the same generating function as the cumulants studied in this paper, and our analysis does not apply to them.}
Their expressions are derived by expanding the generating function of cumulants~\cite{Taghavi:2020gcy}, which depends on complex variables~\cite{Mehrabpour:2020wlu} $\lambda$ and $\mu$ and their complex conjugates  $\lambda^*$ and $\mu^*$:

\begin{align}
G_{\rm lab.}(\lambda,\mu) &\equiv
\ln\left\langle\exp\left(\lambda^*V_2+\lambda V_2^*+\mu^* V_3 +\mu V_3^*\right)\right\rangle\nonumber\\
&=\sum_{m,p,q,r\ge 0} \frac{(\lambda^*)^{m}\lambda^p(\mu^*)^{q}\mu^r}{m!\,p! \,q!\, r!} c^{\rm lab.}_{mpqr}.
\label{labcumulants}
\end{align}
In the first line, angular brackets  in the right-hand side denote an average over events in a centrality class. 
Azimuthal symmetry requires that the generating function is symmetric under $(\lambda,\mu) \to (\lambda e^{2i\varphi},\mu e^{3i\varphi})$.
This in turn implies that  $c^{\rm lab.}_{mpqr}$ may differ from $0$ only if $2(m-p)+3(q-r)=0$. 
A sufficient condition is to have $m=p$ {\it and\/} $q=r$. 
The corresponding cumulants are precisely the mixed harmonic cumulants:
\begin{equation}
MHC(v_2^{2m},v_3^{2q})
\equiv  c^{\rm lab.}_{mmqq}.
\label{defMHC}
\end{equation}
There are other non-vanishing mixed cumulants, such as 
$c^{\rm lab.}_{3002}$.  
They involve angular correlations between $V_2$ and $V_3$~\cite{Bhalerao:2011yg}, also referred to as event-plane correlations~\cite{ATLAS:2014ndd}, and will not be studied here. 

$MHC(v_2^{2m},v_3^{2q})$ is a cumulant of order $2(m+q)$. 
If $q=0$ or $m=0$, it corresponds to the cumulant of a single harmonic $n=2$ or $n=3$, usually denoted by $c_n\{2k\}$~\cite{Borghini:2001vi,ATLAS:2019peb} (see Appendix~\ref{s:v2cumulants}): 
\begin{align}
MHC(v_2^{2m},v_3^0)&=c_2\{2m\}\nonumber\\
MHC(v_2^{0},v_3^{2q})&=c_3\{2q\}.
\label{notmixed}
\end{align} 
The mixed cumulants are those for which both $m$ and $q$ are positive. 
The lowest-order mixed cumulant corresponds to $m=q=1$. 
Expanding the left-hand side of Eq.~(\ref{defMHC}) to order $\lambda\lambda^*\mu\mu^*$, one obtains its expression in terms of moments: 
\begin{equation}
MHC(v_2^2, v_3^2) =  \langle  v_2^2   v_3^2  \rangle-    \langle  v_2^2\rangle  \langle   v_3^2  \rangle.
\label{MHC22}
\end{equation}
It was measured by ALICE in 2016 in Ref.~\cite{ALICE:2016kpq}, where it was named $SC(3,2)$. 
Higher-order cumulants with $(m,q)=(2,1),(3,1),(1,2),(2,2),(1,3)$ were subsequently measured in Ref.~\cite{ALICE:2021adw}, where their expressions in terms of moments are provided. 
Deriving these expressions is straightforward using Eq.~(\ref{defMHC}). 
We do not repeat them because they are lengthy and will not be needed in this work.\footnote{  \label{phases}
Note that the six MHC cumulants analyzed by ALICE and CMS only involve the magnitudes, not the phases of $V_2$ and $V_3$. 
The phases only enter at higher orders: The expression of  $MHC(v_2^6, v_3^4)$ (a cumulant of order 10) in terms of moments   involves a term proportional to $\langle V_2^3(V_3^{*})^2\rangle\langle (V_2^*)^3V_3^2\rangle$.}

The cumulants are usually normalized as follows: 
\begin{equation}
nMHC(v_2^{2m},v_3^{2q})\equiv\frac{MHC(v_2^{2m},v_3^{2q})}{\langle v_2^{2m}\rangle \langle v_3^{2q}\rangle}. 
\label{defnMHC}
\end{equation}
This normalization suppresses the sensitivity to kinematic cuts: 
If $v_n$ is proportional to the initial anisotropy $\varepsilon_n$, the dependence on kinematic cuts can only be in the proportionality factor (the linear response coefficient) which cancels in the  ratio (\ref{defnMHC}). 
The normalized coefficient provides an intuitive dimensionless measure of the strength of the correlation. 
A value of order unity implies a strong correlation between $v_2$ and $v_3$.

\begin{figure}[ht]
   \includegraphics[width=\linewidth]{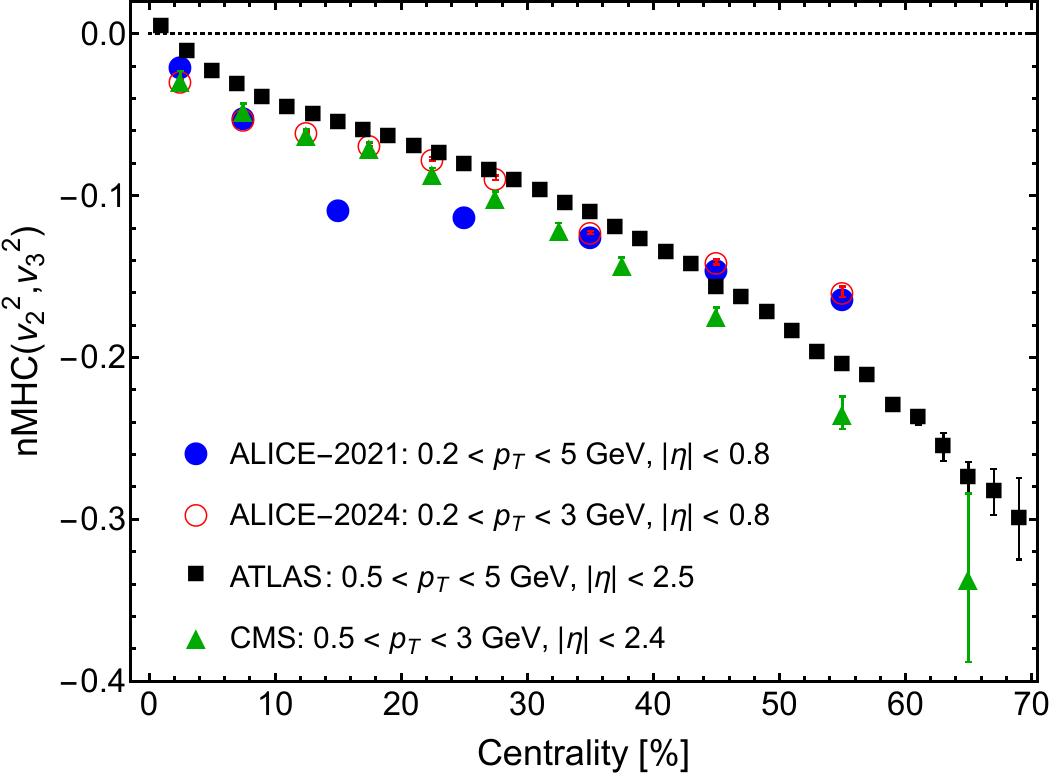}
   \caption{
   Centrality dependence of $nMHC(v_2^2,v_3^2)$ in Pb-Pb collisions at 5.02 TeV:
   Comparison between  2021~\cite{ALICE:2021adw} and 2024~\cite{ALICE:2024nqd} ALICE results, ATLAS~\cite{ATLAS:2019peb} and CMS~\cite{CMS:2025opi}. 
   Note that this quantity is denoted by $NSC(3,2)$ in the 2024 ALICE paper, and by $nsc_{2,3}\{4\}$ in the ATLAS paper. 
}      
        \label{fig:ALICEvsATLAS}
\end{figure}

\subsection{Experimental data}
\label{s:expdata}

There have been several analyses of the lowest-order cumulant $nMHC(v_2^2,v_3^2)$ in Pb+Pb collisions at the LHC, which we first compare. 
We then compare ALICE and CMS results for the higher-order cumulants. 

The normalization (\ref{defnMHC}) facilitates the comparison between experiments, because of the reduced sensitivity to kinematic cuts. 
Fig.~\ref{fig:ALICEvsATLAS} displays four sets of experimental results for $nMHC(v_2^2,v_3^2)$ obtained by the ALICE Collaboration in 2021~\cite{ALICE:2021adw} and 2024~\cite{ALICE:2024nqd},  by ATLAS~\cite{ATLAS:2019peb}, and recently by CMS~\cite{CMS:2025opi}.
One notices that the 2024 ALICE results differ significantly from the 2021 results between 10\% and 30\%  centrality. 
The 2024 analysis is done in narrower bins (5\% as opposed to 10\%). 
The sensitivity of this specific observable to the width of the centrality bins has already been pointed out~\cite{Gardim:2016nrr}, and narrower bins are preferred. 
Results from ATLAS, CMS, and ALICE 2024 are in fair agreement. 
The origin of the residual discrepancies is unknown. 
It could be due to the kinematic cuts, which are different for the four analyses. 
Note that the dependence on $p_T$ cuts cannot be explained within the simple picture of linear hydrodynamic response. 
The negative sign of $nMHC(v_2^2,v_3^2)$ and the centrality dependence can be ascribed to non-Gaussian fluctuations in the initial state~\cite{Alqahtani:2025wan}, as will be recalled in Sec.~\ref{s:expvsintrinsic}. 

\begin{figure}[ht]
   \includegraphics[width=\linewidth]{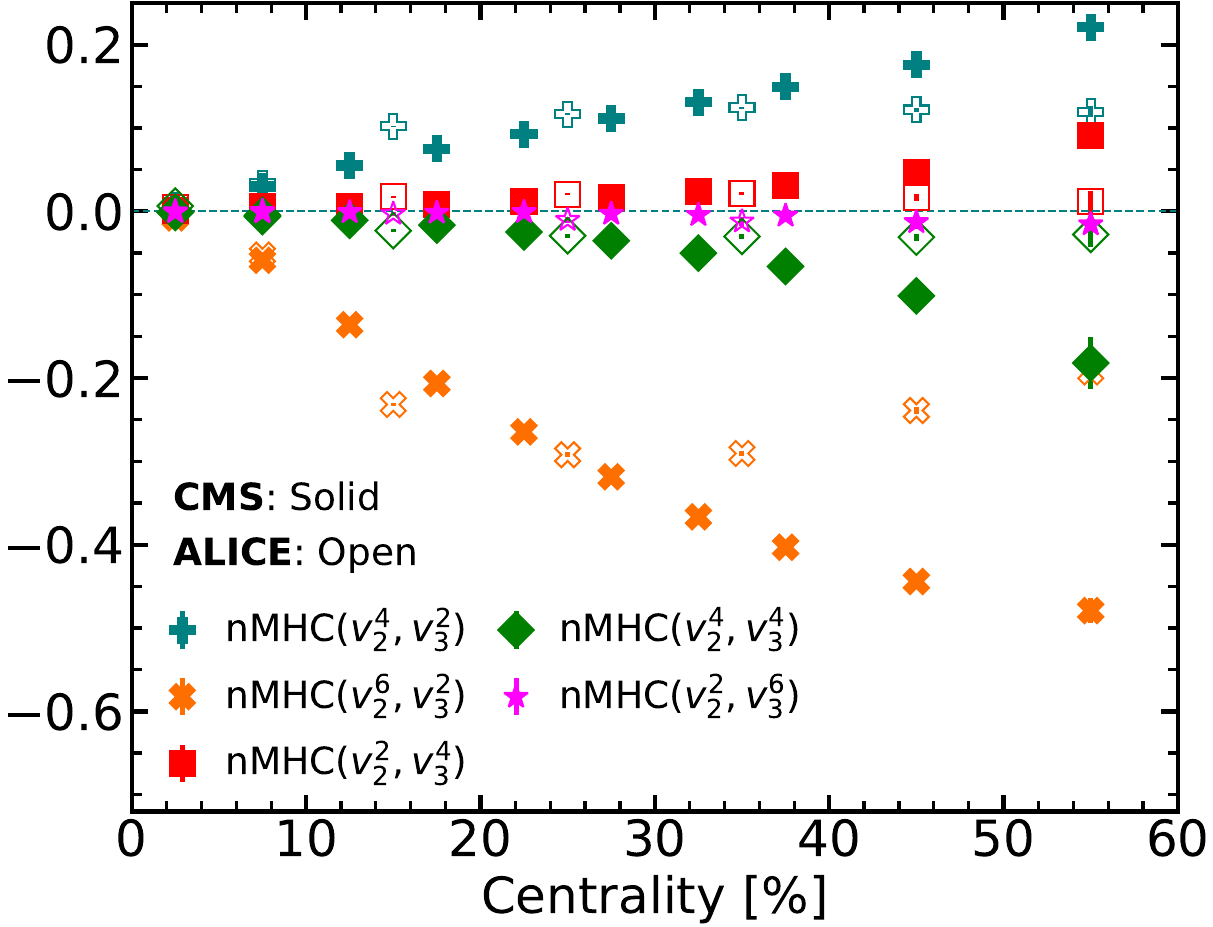}
    \caption{Comparison between ALICE~\cite{ALICE:2021adw} and CMS \cite{CMS:2025opi} results for higher-order normalized mixed harmonic cumulants.}      
        \label{fig:ALICECMSMHC}
\end{figure}

In this paper, we study higher-order cumulants which were measured in by ALICE in 2021~\cite{ALICE:2021adw} and more recently by CMS~\cite{CMS:2025opi}. 
The two sets of results are compared in Fig.~\ref{fig:ALICECMSMHC} as a function of the collision centrality. 
The normalized cumulants are, in absolute magnitude, smaller than unity, which means that correlations between $v_2$ and $v_3$ are not strong.\footnote{By contrast, $nMHC(v_2^{2},v_4^{2})$ reaches $1$ in peripheral collisions~\cite{ATLAS:2019peb,ALICE:2021adw}, which is a natural consequence of the non-linear coupling between $v_2$ and $v_4$~\cite{Teaney:2012ke}.}
The sign of $nMHC(v_2^{2m},v_3^{2q})$  alternates as $m$ or $q$ increases. 
In absolute magnitude, it increases somewhat with $m$, and decreases strongly with $q$. 
These features will be explained in Sec.~\ref{s:expvsintrinsic}.  

As already noticed for the lowest-order cumulant in Fig.~\ref{fig:ALICEvsATLAS}, there are sizable differences between ALICE and CMS results. 
CMS uses 5\% centrality bins between 10\% and 40\% centrality, as opposed to 10\% bins for ALICE, which may explain some of the discrepancies in this range. 
The CMS cumulants increase monotonically  with the centrality fraction in absolute magnitude in agreement with hydrodynamics predictions shown in~\cite{ALICE:2021adw,CMS:2025opi}, while the ALICE cumulants reach a maximum and then decrease, and we do not know the origin of this difference. 

\begin{figure}[ht]
   \includegraphics[width=\linewidth]{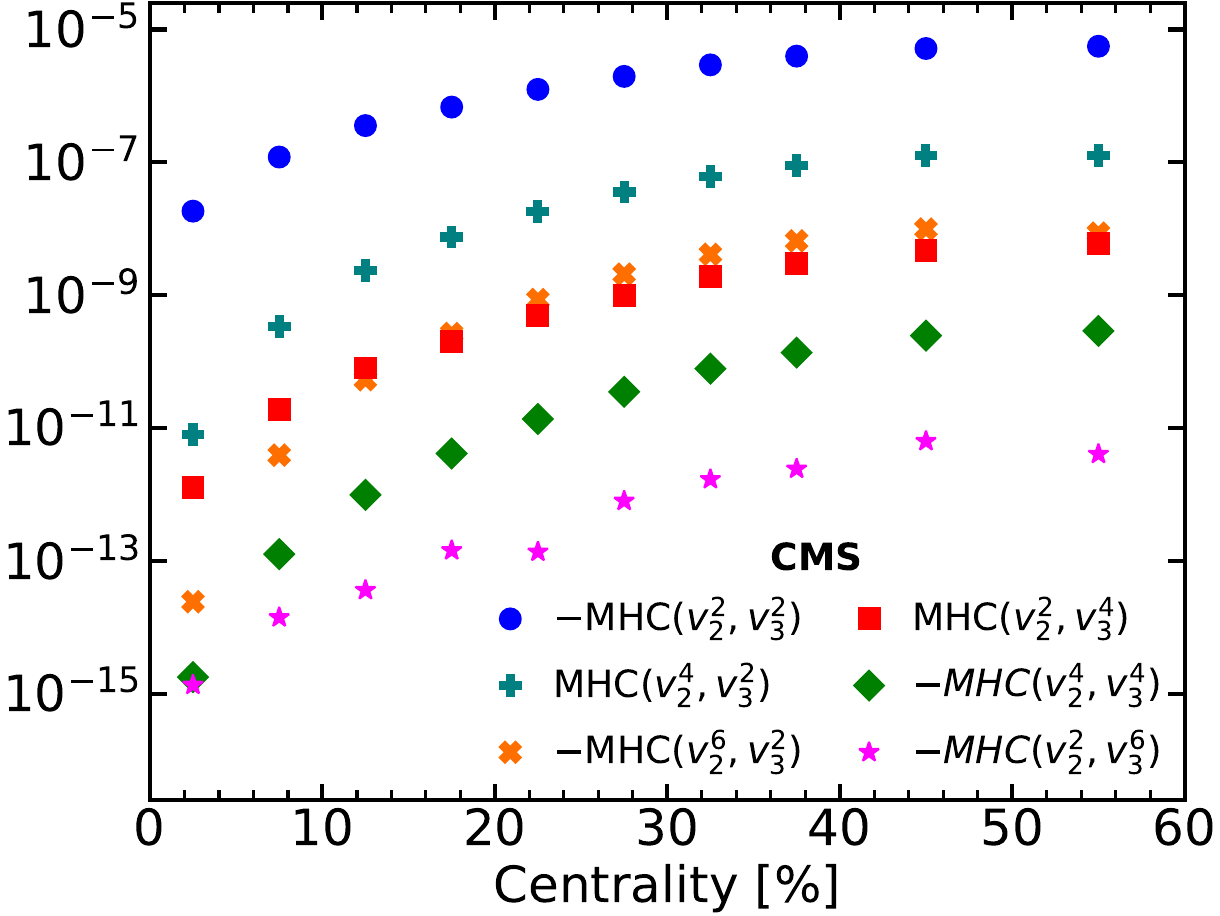}
    \caption{
    Mixed harmonic cumulants as a function of centrality for Pb+Pb collisions at 5.02~TeV per nucleon pair, using CMS data~\cite{CMS:2025opi} for the normalized cumulants and evaluating  the  denominator of (\ref{defnMHC}) with formulas in Appendix~\ref{s:v2cumulants}.
}      
        \label{fig:CMSMHC}
\end{figure}

For our analysis, we will need the un-normalized cumulants $MHC(v_2^{2m},v_3^{2q})$. 
In order to obtain them from the normalized cumulants, we evaluate the moments appearing in the denominator of Eq.~(\ref{defnMHC}) using standard formulas which are recalled in Appendix~\ref{s:v2cumulants}. 
For the sake of illustration, CMS results are displayed in  Fig.~\ref{fig:CMSMHC} as a function of the collision centrality. 
One sees that they span ten orders of magnitude, a point to which we come back in Sec.~\ref{s:expvsintrinsic}.

\section{Cumulants of flow fluctuations in the intrinsic frame}
\label{s:intrinsiccumulants}

\begin{figure}[ht]
   \includegraphics[width=\linewidth]{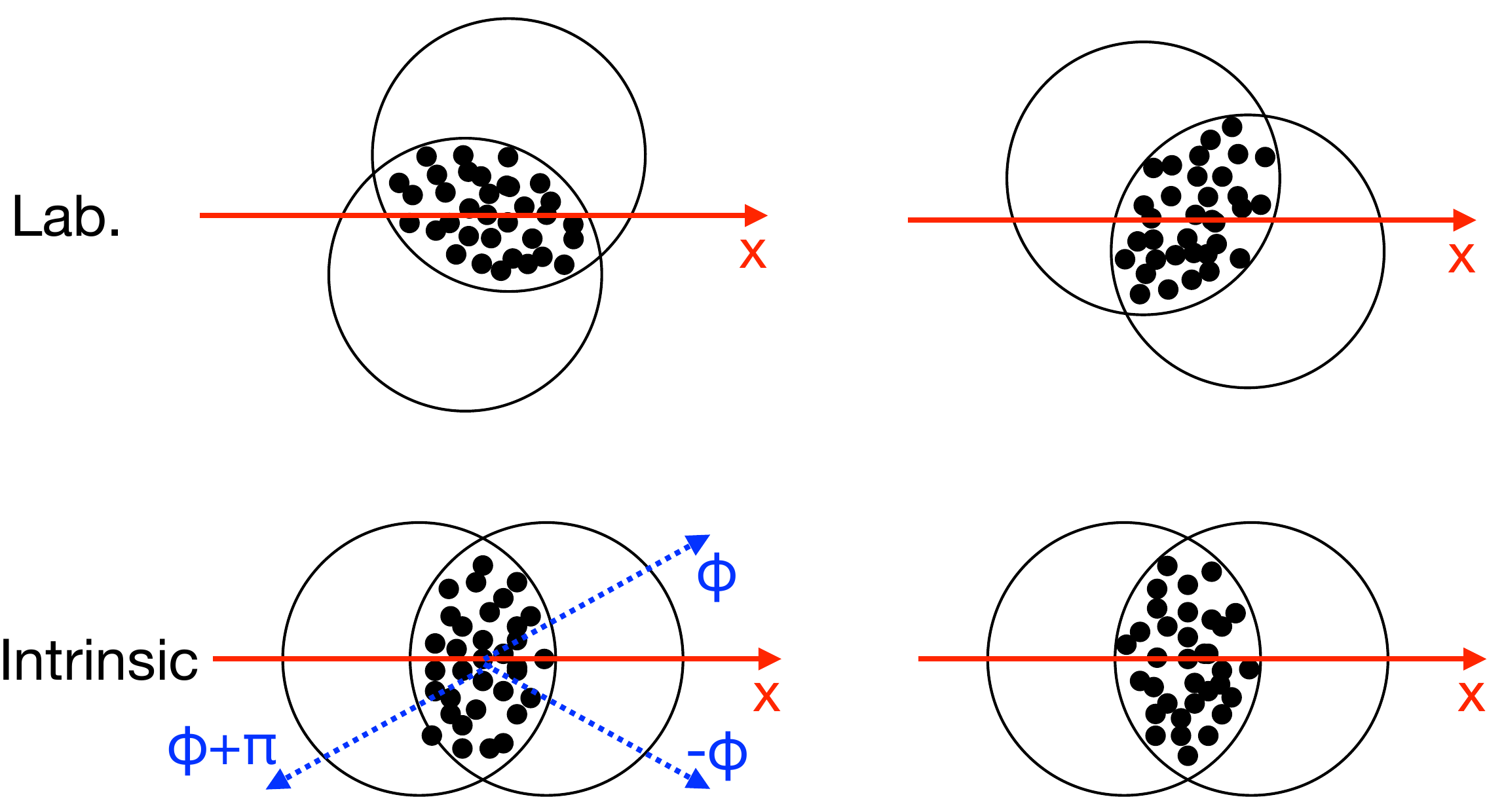}
   \caption{Schematic representation of two collision events with the same impact parameter seen in the laboratory frame and in the intrinsic frame. The dots in the overlap area correspond to the positions of participants nucleons at the time of impact, which takes a snapshot of the nuclear wavefunction. 
}      
        \label{fig:intrinsic}
\end{figure}
We now carry out a thought experiment where all events are aligned in such a way that their impact parameter is along the $x$ axis.\footnote{The $(x,z)$ plane is the reaction plane, and the $x$ axis is often called the reaction plane by a slight abuse of language.} 
We call this the ``intrinsic frame''~\cite{Alqahtani:2024ejg,Roubertie:2025qps}, as illustrated in Fig.~\ref{fig:intrinsic}. 
More detailed information is available in the intrinsic frame than in the laboratory frame, in which the direction of impact parameter is unknown. 
Only a small subset of this information can be reconstructed from experimental data, as will be illustrated below. 
But this limited information sheds light on the physics underlying the measured cumulants. 

\subsection{Definitions and general properties}
\label{s:generalproperties}

The cumulants of $V_2$ and $V_3$ in the intrinsic frame are defined by a generating function which is formally identical to Eq.~(\ref{labcumulants}), the only difference being that one averages over events with the same reaction plane:   
\begin{align}
G_{\rm intr.}(\lambda,\mu) &\equiv
\ln\left\langle\exp\left(\lambda^*V_2+\lambda V_2^*+\mu^* V_3 +\mu V_3^*\right)\right\rangle\nonumber\\
&=\sum_{m,p,q,r\ge 0} \frac{(\lambda^*)^{m}\lambda^p(\mu^*)^{q}\mu^r}{m!\,p! \,q!\, r!} c_{mpqr}.
\label{intrinsiccumulants}
\end{align}
For central collisions with $b=0$, the generating functions (\ref{labcumulants}) and (\ref{intrinsiccumulants}) are identical. 
Intrinsic cumulants coincide with experimental cumulants, $c_{mpqr}=c^{\rm lab.}_{mpqr}$, and the only non-vanishing cumulants are those with $2(m-p)+3(q-r)=0$. 

For non-central collisions, azimuthal symmetry is lost since the $x$ axis is the reaction plane. 
Two weaker symmetries remain: 
Symmetry with respect to the reaction plane, $\varphi\to -\varphi$ (Fig.~\ref{fig:intrinsic}), implies that $c_{mpqr}$ is real and $c_{pmrq}=c_{mpqr}$. 
Symmetry between target and projectile, $\varphi\to \varphi+\pi$, further implies that $c_{mpqr}$ vanishes for odd  $q+r$.

The essential difference between the intrinsic frame and the laboratory frame is that there is now a non-trivial cumulant of order $1$ allowed by symmetry, namely, the mean elliptic flow in the reaction plane. 
We denote it by $\bar V_2$: 
\begin{equation} 
\bar V_2 \equiv c_{1000}=c_{0100}.\label{meanv2}
\end{equation}
There are four cumulants of order $2$ allowed by symmetry: $c_{1100}$, $c_{0011}$, $c_{0200}=c_{2000}$ and $c_{0002}=c_{0020}$. 
The first two are the largest and correspond to the width of $v_n$ fluctuations, $\sigma_{v_n}$~\cite{Voloshin:2007pc}:\footnote{The other two cumulants $c_{0200}$ and $c_{0002}$ measure the asymmetry of $v_n$ fluctuations, and are much smaller, as will be shown in Sec.~\ref{s:powercounting}.}
\begin{align} 
\sigma_{v_2}^2 &\equiv c_{1100}\nonumber\\
\sigma_{v_3}^2 &\equiv c_{0011}.\label{variances}
\end{align}
If fluctuations in the intrinsic frame are Gaussian~\cite{Voloshin:2007pc}, all cumulants vanish beyond order $2$. 
Now, the cumulants of order $1$ and $2$ listed above do not mix $V_2$ and $V_3$. 
Therefore, $V_2$ and $V_3$ are uncorrelated in the  Gaussian limit~\cite{Alqahtani:2025wan}, and  the correlations studied in this paper are all driven by non-Gaussian properties of the distribution. 
In order to model these non-Gaussianities, we need to take into account cumulants of order $3$ and higher.\footnote{
Non-Gaussian fluctuations are also responsible for the non-zero $v_3\{4\}$ seen in Pb+Pb~\cite{ALICE:2011ab} and Xe+Xe~\cite{CMS:2019cyz,CMS:2025xtt} collisions , the hierarchy of higher-order cumulants of $v_2$ ($v_2\{4\}$, $v_2\{6\}$, $v_2\{8\}$) in p+Pb collisions~\cite{CMS:2015yux,CMS:2019wiy}, as well as the splitting between $v_2\{4\}$ and $v_2\{6\}$ in mid-central Pb+Pb collisions~\cite{Giacalone:2016eyu,CMS:2017glf,ALICE:2018rtz,CMS:2023bvg}.}
In Sec.~\ref{s:powercounting}, we discuss the orders of magnitude of these higher-order cumulants and, more specifically, how they depend on the size and shape of the quark-gluon plasma created in the early stages of the collision. 

We have seen that experimental and intrinsic cumulants coincide if $b=0$, which implies $MHC(v_2^{2m},v_3^{2q})=c_{mmqq}$ using Eq.~(\ref{defMHC}). 
For $b\not= 0$, on the other hand, the expression of $MHC(v_2^{2m},v_3^{2q})$ as a function of intrinsic cumulants involves additional terms, which will be derived in Sec.~\ref{s:expvsintrinsic}. 

\subsection{Power-counting scheme}
\label{s:powercounting}

We now derive general scaling laws that apply to the cumulants in the intrinsic frame. 
For simplicity, we neglect centrality fluctuations, and we assume that all events in a centrality class have the same impact parameter. 
Impact parameter fluctuations have large effects for the most central collisions~\cite{Alqahtani:2024ejg}, but we have found that their effect on correlations between $v_2$ and $v_3$ is modest beyond 5\% centrality~\cite{Alqahtani:2025wan}, provided that the analysis uses narrow centrality bins~\cite{ATLAS:2019peb}. 
The ALICE experiment uses wide centrality bins and this has a significant effect on the results, as discussed in Sec.~\ref{s:expdata}, but this can easily be improved in future analyses. 

If the impact parameter is constant, different events in the intrinsic frame differ only by local fluctuations in the initial density profile, which are of quantum origin, as illustrated in Fig.~\ref{fig:intrinsic}. 
The resulting fluctuations of the complex anisotropies $\varepsilon_2$ and $\varepsilon_3$~\cite{Qiu:2011iv}, which generate $V_2$ and $V_3$ through linear hydrodynamic response, originate from a large number $N$ of independent local fluctuations.\footnote{ 
$N$ is typically of the same order as the number of participant nucleons~\cite{Miller:2007ri}.}
For independent fluctuations, a cumulant of order $k$ varies with $N$ like $N^{1-k}$, where the order is the sum of indices $k=m+p+q+r$. 
The variances in Eq.~(\ref{variances}) are cumulants of order $2$, and are therefore of order $1/N$. 
We denote by  ${\cal F}$ the typical magnitude of event-by-event flow fluctuations, as measured by $\sigma_{v_2}$ and $\sigma_{v_3}$, which are comparable~\cite{Roubertie:2025qps}, that is, ${\cal F} \propto 1/\sqrt{N}$. 
Then, a cumulant of order $k$ is of order $N^{1-k}\propto {\cal F}^{2k-2}$.  

We now discuss how cumulants depend on the shape of the system, measured by its mean deformation. 
We assume that azimuthal symmetry is mildly broken, $\bar V_2\ll 1$. 
The order of magnitude of $c_{mpqr}$ depends on the shape through the order of the Fourier coefficient it corresponds to, $2n=|2(m-p)+3(q-r)|$. 
For mildly-broken azimuthal symmetry, a Fourier coefficient of order $2n$ is suppressed by a factor $(\bar V_2)^n$. 
For instance, the asymmetries of $v_2$ and $v_3$ fluctuations,  $c_{0200}$ and  $c_{0002}$, are suppressed by factors $\bar V_2^2$ and $\bar V_2^3$ with respect to $c_{1100}$ and  $c_{0011}$.  

Putting together the dependences on the size and shape, we generally expect
\begin{equation}
c_{mpqr} \sim {\cal O}\left({\cal F}^{2(m+p+q+r-1)} \bar V_2^{\left|m-p+\frac{3}{2}(q-r)\right|}\right).\label{powercounting}
\end{equation}
This gives for instance $c_{1111}\sim {\cal O}\left({\cal F}^6\right)$. 
Therefore, in central collisions, where $MHC(v_2^{2},v_3^{2})=c_{1111}$, each of the moments in the right-hand side of Eq.~(\ref{MHC22}) is of order ${\cal F}^4$, while the difference is of order ${\cal F}^6$, i.e., much smaller. 
This systematic expansion scheme will allow us to single out the dominant contributions to each of the MHCs. 

\section{Relations between experimental cumulants and intrinsic cumulants}
\label{s:expvsintrinsic}

We now relate the two sets of cumulants defined in Secs.~\ref{s:expcumulants} and \ref{s:intrinsiccumulants}. 
In the laboratory frame, the orientation of impact parameter is uniformly distributed. 
Therefore, moments of $V_2$, $V_3$,  $V_2^*$, $V_3^*$  in the laboratory frame are obtained from the corresponding moments in the intrinsic frame by averaging over all possible orientations of impact parameter.  
This implies the following relation between the generating functions (\ref{defMHC}) and (\ref{intrinsiccumulants})~\cite{Abbasi:2017ajp}: 
\begin{equation}
G_{\rm lab.}(\lambda,\mu)=\ln\left( \int_{0}^{2\pi} \frac{d\varphi}{2\pi} \exp\left( G_{\rm intr.}(\lambda e^{2i\varphi},\mu e^{3i\varphi})\right)\right). 
\label{labvsintr}
\end{equation}
Expanding to order $\lambda\lambda^*\mu\mu^*$, one obtains the following exact expression of the lowest-order mixed cumulant (\ref{MHC22}) as a function of the intrinsic cumulants~\cite{Alqahtani:2025wan}: 
\begin{align}
MHC(v_2^2, v_3^2) &= c_{1000}c_{0111}+c_{0100}c_{1011}+c_{1111}\nonumber\\
&= 2\bar V_2 c_{0111}+c_{1111}
\nonumber\\
&\sim \bar V_2^2{\cal F}^{4}+{\cal F}^{6}. 
\label{MHC22int}
\end{align}
where we have introduced $\bar V_2$ defined by Eq.~(\ref{meanv2}), used the symmetry $c_{mpqr}=c_{pmrq}$, and finally Eq.~(\ref{powercounting}) to evaluate the orders of magnitude of the terms. 
Only the second term contributes for central collisions where $\bar V_2=0$. 
As the centrality fraction increases, the first term gradually becomes the dominant term~\cite{Alqahtani:2025wan}. 

The higher-order cumulants measured by ALICE and CMS~\cite{ALICE:2021adw,CMS:2025opi} can also be expressed exactly in terms of intrinsic cumulants. 
In order to simplify expressions, we only retain the largest terms. 
More specifically, we assume that ${\cal F}$ and $\bar V_2$ are generically of the same order of magnitude, which means that we consider that both terms in Eq.~(\ref{MHC22int}) are of the same order. 
We then truncate expressions to leading order in the small parameter ${\cal F}\sim\bar V_2$. 
We obtain: 
\begin{align}
MHC(v_2^4, v_3^2) &= -4\,\bar V_2^3\, c_{0111}\sim \bar V_2^4{\cal F}^4\nonumber\\
MHC(v_2^6, v_3^2) &= 24\,\bar V_2^5 \, c_{0111} \sim \bar V_2^6{\cal F}^4\nonumber\\
MHC(v_2^2, v_3^4) &= 2 (2\, c_{0111}^2+\bar V_2\, c_{0122})+ c_{1122}\nonumber\\
&\sim \bar V_2^2{\cal F}^8+{\cal F}^{10}\nonumber\\
MHC(v_2^4, v_3^4) &= -4 \bar V_2^2 (6 c_{0111}^2+ \bar V_2 c_{0122}) \sim \bar V_2^4{\cal F}^8\nonumber\\
MHC(v_2^2, v_3^6) &= 2(9 c_{0111}c_{0122}+\bar V_2c_{0133})+ c_{1133}\nonumber\\
&\sim \bar V_2^2{\cal F}^{12}+{\cal F}^{14}.\label{leadingorder}
\end{align}
If $\bar V_2\sim {\cal F}$, they are respectively of order ${\cal F}^8$,  ${\cal F}^{10}$, ${\cal F}^{10}$, ${\cal F}^{12}$, ${\cal F}^{14}$. 
More generally, $MHC(v_2^{2m}, v_3^{2q})$ is of order ${\cal F}^{2m+4q}$. 
These orders of magnitude are reflected in the hierarchy observed in Fig.~\ref{fig:CMSMHC}. 
In particular, they explain why $MHC(v_2^6, v_3^2)$ and $MHC(v_2^2, v_3^4)$ are of comparable magnitude, despite being cumulants of different orders  (8 and 6 respectively). 

Since $v_2$ and $v_3$ are both of order ${\cal F}$, the normalized symmetric cumulant (\ref{defnMHC}) is of order ${\cal F}^{2q}$. 
This explains why the magnitude of $nMHC(v_2^{2m},v_3^{2q})$ decreases strongly as $q$ increases, as seen in Fig.~\ref{fig:ALICECMSMHC}. 

We finally provide leading-order expressions for the cumulants of order 10, which have not yet been analyzed: 
\begin{align}
MHC(v_2^8, v_3^2) =&-264\,\bar V_2^7 \,c_{0111}\sim \bar V_2^8{\cal F}^{4}\nonumber\\
MHC(v_2^6, v_3^4) =&\, 24\, \bar V_2^4 (10\,  c_{0111}^2+ \bar V_2\,  c_{0122})\sim \bar V_2^6{\cal F}^8\nonumber\\
MHC(v_2^4, v_3^6) =&-144\, \bar V_2\,  c_{0111}^3-108\bar V_2^2 c_{0111}c_{0122}\nonumber\\&-4\bar V_2^3c_{0133}\sim \bar V_2^4{\cal F}^{12}\nonumber\\
MHC(v_2^2, v_3^8) =&36\,c_{0122}^2+32  c_{0111}c_{0133}+4\bar V_2c_{0144}\nonumber\\&+c_{1144}\sim \bar V_2^2{\cal F}^{16}+{\cal F}^{18} .
\label{leadingorder10}
\end{align}
The first three could easily be measured (note however that the second involves event-plane correlations, see footnote \ref{phases}), while the last is likely too small. 

A first comment on the expressions (\ref{MHC22int}), (\ref{leadingorder}) and (\ref{leadingorder10}) is that they only involve the mean elliptic flow in the reaction plane, $\bar V_2$, and mixed cumulants of order $\ge 3$, which quantify non-Gaussian fluctuations.
Thus the correlations between $v_2$ and $v_3$ originate from non-Gaussian fluctuations in the initial state~\cite{Alqahtani:2025wan}. 

A second remark is that for a given value of $q$, $MHC(v_2^2, v_3^{2q})$ contains more terms than higher-order cumulants $MHC(v_2^{2m}, v_3^{2q})$ with $m\ge 2$, whose expressions are somewhat simpler. 
More specifically, the expressions for $m=1$ contain the cumulant $c_{11qq}$, which no longer appears for $m\ge 2$. 
For a given value of $q$, each of the remaining terms is multiplied by a factor proportional to  $\bar V_2^2$ as $m$ increases. 
This can be related to the well-known fact that cumulants of $v_2$ of order 4 and higher only depend on $\bar V_2$ to leading order in ${\cal F}$, as recalled in Appendix~\ref{s:v2cumulants}.
In addition, there are overall numerical factors which increase rapidly as a function of $m$ ($2$, $-4$, $24$, $-264$ for $m=1,2,3,4$). 
These numerical factors explain why the normalized cumulant $nMHC(v_2^{2m}, v_3^{2q})$ changes sign and increases in absolute magnitude as $m$ increases, as seen in Fig.~\ref{fig:ALICECMSMHC}. 
 
\section{Relations between cumulants of different orders}
\label{s:ratios}

\begin{figure}[ht]
   \includegraphics[width=\linewidth]{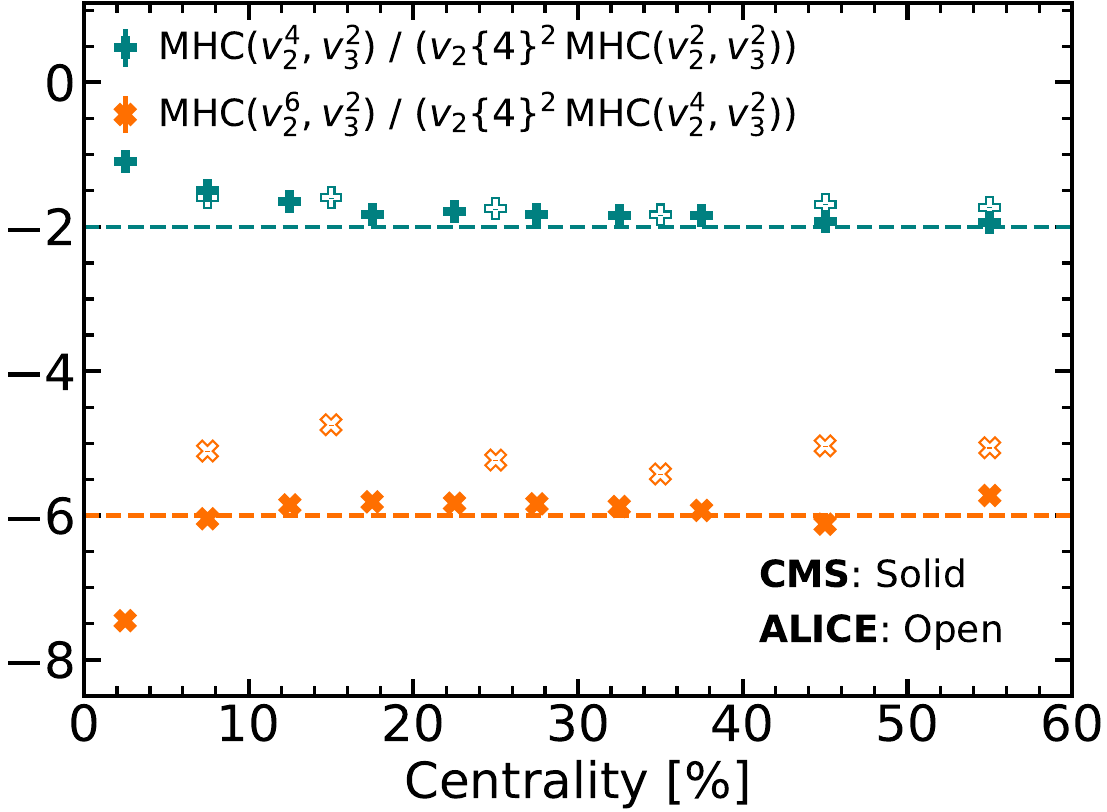}
    \caption{Ratios in Eqs.~(\ref{ratios0}) and (\ref{ratios1})  as a function of the collision centrality in Pb+Pb collisions at 5.02~TeV per nucleon pair.   
    Open symbols are ALICE data, where the mixed cumulants are taken from Ref.~\cite{ALICE:2021adw} and $v_2\{4\}$ from Ref.~\cite{ALICE:2016ccg}. 
    Closed symbols are CMS data~\cite{CMS:2025opi}.
    Horizontal lines are our theory predictions, Eqs.~(\ref{ratios0}) and (\ref{ratios1}). 
}      
        \label{fig:ratios}
\end{figure}

Eqs.~(\ref{MHC22int}), (\ref{leadingorder}) and (\ref{leadingorder10}) show that to leading order, the experimental cumulants only depend on a small number of intrinsic cumulants. 
First of all, they involve the mean elliptic flow in the reaction plane, which can be obtained from data using the approximation $\bar V_2\approx  v_2\{4\}$ (App.~\ref{s:v2cumulants}). 
ALICE only measures $v_2\{4\}$ above 5\% centrality~\cite{ALICE:2016ccg} so that we exclude the most central bin 0-5\%.

The remaining intrinsic cumulants can be eliminated by combining the information from several measured cumulants. 
Taking the ratio between the first two lines of Eq.~(\ref{leadingorder}), for instance, one obtains: 
\begin{equation}
\frac{MHC(v_2^6, v_3^2)}{v_2\{4\}^2 MHC(v_2^4, v_3^2)}= -6.\label{ratios0}
\end{equation}
Fig.~\ref{fig:ratios} shows that CMS data agree very well with this equation, except for the most central bin, where impact parameter fluctuations may have a sizable effect. 
Agreement with ALICE data is poorer. 

Similarly, using Eqs.~(\ref{leadingorder}) and (\ref{leadingorder10}), we predict: 
\begin{equation}
\frac{MHC(v_2^8, v_3^2)}{v_2\{4\}^2 MHC(v_2^6, v_3^2)}= -11.\label{ratios2}
\end{equation}
This could easily be checked experimentally, as increasing the order in $v_2$ does not significantly increase errors. 

Equations~(\ref{ratios0}) and (\ref{ratios2}) are rigorous mathematical results to leading order in  ${\cal F}$. 
They generalize the well-known identities $v_2\{4\}=v_2\{6\}=v_2\{8\}$ (Appendix~\ref{s:v2cumulants}) to mixed cumulants. 
We therefore expect that their accuracy is comparable, at the percent level in Pb+Pb collisions at LHC energies~\cite{CMS:2017glf,ALICE:2018rtz,CMS:2023bvg}, provided that the analysis is carried out in fine centrality bins.\footnote{The intrinsic centrality resolution at the LHC is of order 2\% \cite{Das:2017ned}. This is the typical order of magnitude of the optimal width of the centrality binning.}

Another similar relation can be derived using Eq.~(\ref{MHC22int}). 
Above 5\% centrality, elliptic flow in the reaction plane dominates over flow fluctuations~\cite{Roubertie:2025qps}, that is, $\bar V_2> {\cal F}$. 
This implies that the kurtosis $c_{1111}$ is smaller than the first term~\cite{Alqahtani:2025wan}. 
Neglecting $c_{1111}$, one obtains:
\begin{equation}
\frac{MHC(v_2^4, v_3^2)}{v_2\{4\}^2 MHC(v_2^2, v_3^2)}\approx -2.\label{ratios1}
\end{equation}
This prediction is also in good agreement with data, as shown in Fig.~\ref{fig:ratios}.
Eqs.~(\ref{ratios0}), (\ref{ratios2}), and (\ref{ratios1}) show that as one increases $m$, the change in  $MHC(v_2^{2m},v_3^2)$ is solely determined by the mean elliptic flow in the reaction plane $\bar V_2\simeq v_2\{4\}$. 
This property also holds for higher orders in $v_3$, which we now study. 
 
\begin{figure}[ht]
   \includegraphics[width=\linewidth]{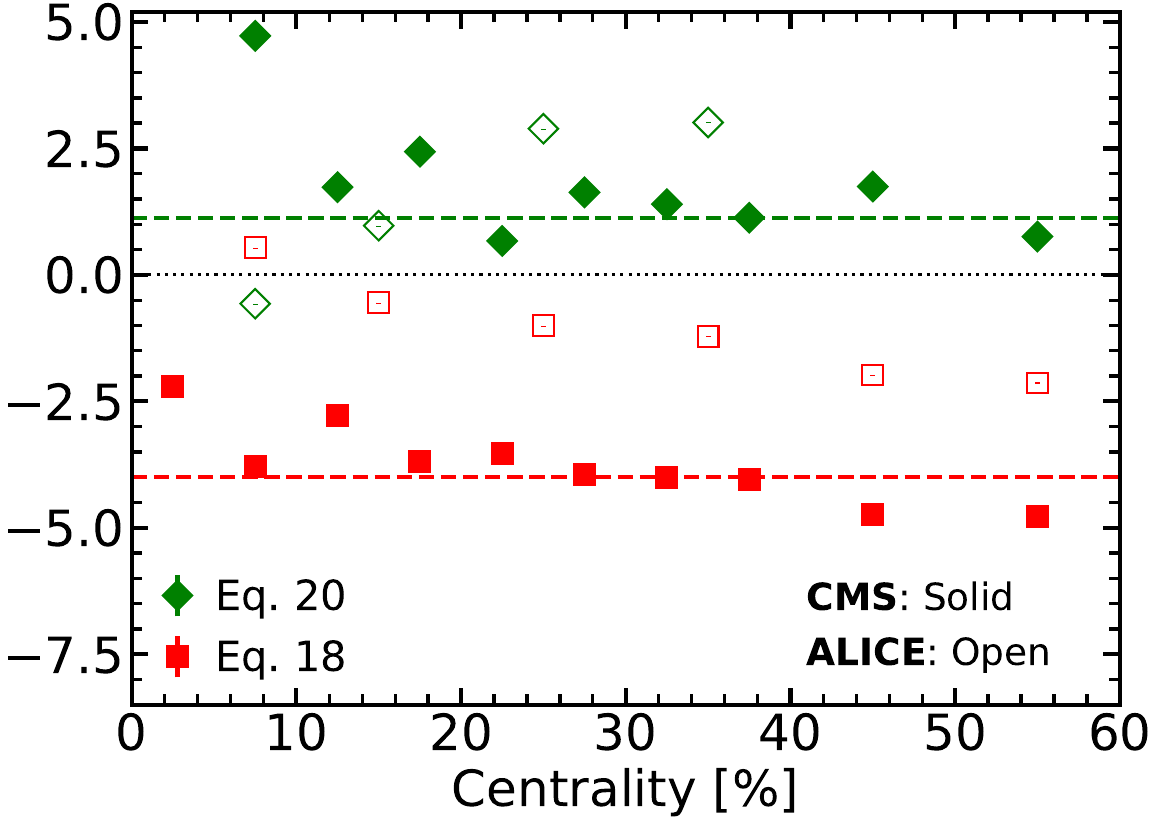}
   \caption{Same as Fig.~\ref{fig:ratios} for the ratios Eqs.~(\ref{ratios3}) and (\ref{ratios4}). 
}      
        \label{fig:ratios2}
\end{figure}
We now move on to the cumulants involving $v_3^4$, third and fourth lines of Eq.~(\ref{leadingorder}). 
They involve the mixed skewness $c_{0111}$, and also new, higher-order cumulants: a mixed ``superskewness''~\cite{CMS:2023bvg}  $c_{0122}$ and superkurtosis $c_{1122}$, which are cumulants of order 5 and 6 respectively. 
The superkurtosis is of order ${\cal F}^{10}$ and we neglect it with respect to the contribution of the superskewness, of order $\bar V_2^2{\cal F}^8$.
In addition, we again neglect the contribution of the kurtosis $c_{1111}$ relative to that of the mixed skewness $c_{0111}$ in Eq.~(\ref{MHC22int}). 
We finally eliminate $c_{0122}$ through a linear combination, and we obtain: 
\begin{equation}
\frac{MHC(v_2^4, v_3^4) +2 v_2\{4\}^2MHC(v_2^2, v_3^4) }{MHC(v_2^2, v_3^2)^2}\approx -4.
\label{ratios3}
\end{equation}
Comparison with experimental data is displayed in Fig.~\ref{fig:ratios2}.
Our prediction agrees well with CMS data, not with ALICE data.  

Another similar relation can be predicted using the cumulant of order $10$ in Eq.~(\ref{leadingorder10}), which is not yet measured: 
\begin{equation}
\frac{v_2\{4\}^2 MHC(v_2^6, v_3^4) +6 v_2\{4\}^4MHC(v_2^4, v_3^4) }{MHC(v_2^4, v_3^2)^2}=6.
\label{ratios5}
\end{equation}
The advantage over Eq.~(\ref{ratios3}) is one need not neglect $c_{1111}$ and $c_{1122}$, which do not enter the leading-order expressions of the cumulants involved in Eq.~(\ref{ratios5}). 
Like Eqs.~(\ref{ratios0}) and (\ref{ratios2}), Eq.~(\ref{ratios5}) is a rigorous leading-order result, and we expect that it should be fairly accurate with a fine centrality binning. 

Finally, we comment on $MHC(v_2^2,v_3^6)$, last line of Eq.~(\ref{leadingorder}). 
In order to relate it to lower-order cumulants, we first  neglect the third term $c_{1133}$, of order ${\cal F}^{14}$, relative to the first two terms which are of order $\bar V_2^2{\cal F}^{12}$. 
We also neglect the second term because it involves the cumulant $c_{0133}$ which does not appear in lower-order cumulants. 
Since it is a priori of the same order as the first term, the relations derived on this basis hold only in order of magnitude.
We finally carry out the same simplifications as in deriving Eqs.~(\ref{ratios1}) and (\ref{ratios3}). 
Simple algebra then gives
\begin{equation}
\frac{v_2\{4\}^4 MHC(v_2^2,v_3^6)}
{MHC(v_2^2,v_3^2)\left(MHC(v_2^4, v_3^4) +6 v_2\{4\}^2MHC(v_2^2, v_3^4) \right)}\approx\frac{9}{8}.
\label{ratios4}
\end{equation} 
This prediction roughly matches CMS data in terms of order of magnitude (Fig.~\ref{fig:ratios2}). 
Here, we do not expect any significant improvement with a finer centrality binning. 

\section{Conclusions}
\label{s:conclusion}

We have derived several analytic relations between cumulants of the joint distribution between $v_2$ and $v_3$ in ultrarelativistic nucleus-nucleus collisions. 
Equations~(\ref{ratios0}), (\ref{ratios1}) and (\ref{ratios3}) agree well with CMS data, and we have made predictions for higher-order cumulants which could be analyzed with existing data, Eqs~(\ref{ratios2}) and (\ref{ratios5}). 
At face value, these relations suggest that cumulants are redundant, in the sense that increasing $m$ (the order of the cumulant in $v_2$) does not bring any additional information. 
But this may not be the end of the story:
Similarly, one could interpret the approximate equalities $v_2\{4\}\simeq v_2\{6\}\simeq v_2\{8\}$ as a hint that $v_2$ cumulants beyond order $4$ are useless, but precision studies have shown that the small violations of these equalities, at the sub-percent level, reveal invaluable information about non-Gaussian fluctuations~\cite{Giacalone:2016eyu,CMS:2017glf,ALICE:2018rtz,CMS:2023bvg,Roubertie:2025qps}.  
In the same way, if the analysis of mixed cumulants is repeated with greater accuracy, one can measure subdominant terms, which we have neglected. 
As an example, the mixed kurtosis, $c_{1111}$, can be obtained by combining the information from $MHC(v_2^2,v_3^2)$ and $MHC(v_2^4,v_3^2)$.
These refinements are left for future work. 

The simplicity of the relations we have derived arises from the general properties of local density fluctuations, which are of quantum origin. 
The classical fluctuations of impact parameter spoil this simplicity~\cite{Roubertie:2025qps,Alqahtani:2025wan}. 
It is therefore essential to work with fine centrality bins. 
The wide centrality bins implemented by ALICE partly explain the differences between data and our predictions. 
The recent CMS data are in much better agreement with our calculation. 
This illustrates that the effect of impact parameter fluctuations is modest. 
We expect that agreement would be further improved with a finer centrality binning. 

It will be important to check whether full hydrodynamic calculations with fluctuating initial conditions, carried out at fixed impact parameter, confirm the validity of our results, which are derived under the simplifying assumption of linear hydrodynamic response. 
Significant deviations from linear response are observed for elliptic flow above 30\% centrality~\cite{Niemi:2015qia,Noronha-Hostler:2015dbi,Hippert:2020kde,Giacalone:2017uqx}, and their effect on mixed harmonic cumulants must be quantitatively assessed. 
The ALICE~\cite{ALICE:2021adw} and CMS~\cite{CMS:2025opi} include comparisons of data with full hydrodynamic calculations, done with the iEBE-VISHNU model~\cite{Zhao:2017yhj} and with the IP-GLASMA+MUSIC+URQMD framework~\cite{McDonald:2017ayb,Gale:2013da,Bass:1998ca}.
These calculations are in fair agreement with experimental results, but have large error bars. 
This can easily be improved. 
Hydrodynamic calculations of $MHC(v_2^{2m},v_3^{2q})$ with $q=1$ should require modest statistics ($q=2$ and $q=3$ are more demanding because the corresponding normalized cumulants are smaller),  provided that one evaluates anisotropic flow directly on the freeze-out surface~\cite{Cooper:1974mv}.
By contrast, both iEBE-VISHNU and IP-GLASMA+MUSIC +URQMD couple the hydrodynamics to a transport calculation~\cite{Bass:1998ca,Petersen:2008dd}. 
This is the state-of-the-art approach when it comes to comparing with experimental data, but it results in  larger statistical errors, induced by the Monte Carlo sampling of hadrons. 
It is likely that effects of the transport phase largely cancel when taking ratios as in Sec.~\ref{s:ratios}. 
It will be also essential to carry out these hydrodynamic  calculations at fixed impact parameter, which is rarely done. 
Impact parameter fluctuations can be studied independently~\cite{Samanta:2023amp,Alqahtani:2024ejg}, and we have argued that their effects should be small above 5-10\% centrality. 

\begin{acknowledgments}
We thank Ante Bilandzic and You Zhou for discussions about the ALICE results, and Aryaa Dattamunsi for discussions about the CMS results.  
M. Alqahtani acknowledges the support of the Research Mobility Program of the French Embassy in Riyadh, which helped to initiate this work. 
\end{acknowledgments}

\appendix
\section{Moments and cumulants of $v_2$}
\label{s:v2cumulants}

The cumulant of order $2k$ of a single harmonic $v_n$, denoted by $c_n\{2k\}$, is a standard observable of heavy-ion collisions. 
It is obtained by setting $m=0$ or $q=0$ in Eq.~(\ref{defMHC}), expanding to order $2k$ in the remaining variable, and using the definition (\ref{notmixed}). 
One obtains the well-known relations
\begin{align}
c_n\{2\}&=v_n\{2\}^2= \langle v_n^2\rangle\nonumber\\
c_n\{4\}&= -v_n\{4\}^4=\langle v_n^4\rangle- 2\langle v_n^2\rangle^2 \nonumber\\
c_n\{6\}&=4v_n\{6\}^6= \langle v_n^6\rangle- 9 \langle v_n^4\rangle\langle v_n^2\rangle +12\langle v_n^2\rangle^3. \label{cn}
\end{align}
Using Eqs.~(\ref{defMHC}), (\ref{notmixed}), (\ref{intrinsiccumulants}) and (\ref{labvsintr}), one shows that for $n=2$, cumulants of order 4 and higher are solely determined by the mean elliptic flow in the reaction plane, $v_2\{4\}=v_2\{6\}=\bar V_2$, up to relative corrections of order ${\cal F}^2$ (at the sub-percent level). 
These corrections originate mostly from the skewness of elliptic flow fluctuations~\cite{Giacalone:2016eyu,CMS:2017glf,ALICE:2018rtz,CMS:2023bvg}, which is $c_{1200}$ in our notation~\cite{Roubertie:2025qps}. 

Inverting Eqs.~(\ref{cn}), one expresses the moments in terms of cumulants:
\begin{align}
\langle v_n^2\rangle &=v_n\{2\}^2 \nonumber\\
\langle v_n^4\rangle &=2 v_n\{2\}^4-v_n\{4\}^4\nonumber\\
 \langle v_n^6\rangle &=6 v_n\{2\}^6 - 9 v_n\{2\}^2 v_n\{4\}^4 +4v_n\{6\}^6.
 \label{momentsversuscumulants}
\end{align}

The cumulants in the right-hand side of Eq.~(\ref{momentsversuscumulants}) are taken from each experiment and used to evaluate the denominators in Eq.~(\ref{defnMHC}). For ALICE results, they are measured with the same centrality bins~\cite{ALICE:2016ccg}. We neglect  $v_3\{4\}$ and $v_3\{6\}$, which are not provided in this reference, and whose contribution is known to be small~\cite{ATLAS:2019peb}. For the 0-5\% centrality bin, $v_2\{4\}$ and $v_2\{6\}$ are undefined (the corresponding cumulants change sign~\cite{STAR:2015mki,ATLAS:2019peb,Alqahtani:2024ejg}) and we set them to zero. On the other hand, for CMS results they are taken from~\cite{CMS:2025opi} with finer centrality bins. We note here that $v_3\{4\}$ is provided in this CMS analysis.

\bibliography{MHC}

\end{document}